\documentclass[aps,prl,twocolumn,groupedaddress]{revtex4}
\usepackage{graphics}

\begin{document}


\title{Statistical mechanics of RNA folding: a lattice approach}


\author{P. Leoni$^{1,}$}
\altaffiliation{Aspirant FWO-Vlaanderen}
\author{C. Vanderzande$^{1,2}$}
\affiliation{$^1$Departement WNI, Limburgs Universitair Centrum, 3590 Diepenbeek, Belgium\\
$^2$Instituut voor Theoretische Fysica, Celestijnenlaan 200D, 3001 Heverlee, Belgium}


\date{\today}

\begin{abstract}
We propose a lattice model for RNA based on a self-interacting two-tolerant
trail. Self-avoidance and elements of tertiary structure are taken into account.
We investigate a simple version of the model in which the
native state of RNA consists of just one hairpin. Using exact
arguments and Monte Carlo simulations we determine the
phase diagram for this case. We show that the denaturation transition
is first order and can either occur directly
or through an intermediate molten phase.
\end{abstract}


\maketitle

\section{I. Introduction}
In recent years, the diversity of roles played by RNA in cellular
processes has become increasingly clear \cite{nature}. RNA is a
heteropolymer which is build from four types of monomers (nucleotides)
\cite{molbiol} and as for proteins, a major issue is the prediction
of the tertiary (folded) structure for a given sequence of nucleotides
\cite{Bustamente}. In some respects, this question is easier to
answer for RNA than for proteins. Firstly, the four nucleotides are
chemically more similar than the twenty amino acids that form
proteins. Secondly, it has been argued \cite{Bustamente} that folding
in RNA is hierarchical in the sense that the energy scales involved in secondary structure elements
are larger than those of tertiary structure. Therefore much attention,
also in the physics literature \cite{Higgs,McCaskill,BH1,Muller,Pagnani,
BH2,BH3,Krzakala}, has been devoted to predicting the secondary structure
of RNA. 

In most of the existing models, self-avoidance has not been fully
taken into account. If one also neglects certain type of monomer-monomer
interactions such as those occuring in pseudoknots and kissing hairpins
\cite{molbiol}, it becomes possible to calculate secondary structures
with a recursive algorithm whose complexity only grows as the third power
of the number of monomers $L$ \cite{McCaskill}. Despite its success,
this approach has several drawbacks. Firstly, the neglect of self-avoidance
leads to unphysical properties for the radius of gyration \cite{Muller},
especially at low temperatures. 
Secondly, the neglect of part of the physically relevant interactions
is often justified {\it a posteriori} from the observation that (for
example) pseudoknots are not very common in real RNA. However, it
would be more attractive to have a model which allows {\it to predict}
the rate of occurence of these structural elements. A first attempt
along these lines, but neglecting self-avoidance, was made in \cite{Orland}.

In this paper, we propose a lattice model for RNA which can take into
account both self-avoidance and most of the physically relevant interactions.
To get a first insight into the properties of the model, we study here
in detail a simplified version, which is a lattice variant of the model
studied in \cite{BH1}. By using rigorous and numerical techniques well-known
from the study of other lattice models of polymers \cite{Carloboek}, we
find that RNA can exist in three phases. In the present work we study
the properties of these phases and of the transitions between them.
In a forthcoming publication, we will investigate questions such as the probability of formation
of pseudoknots.

This paper is organised as follows. In section II we present our lattice
model and show how one can include several aspects of real RNA in
a natural way. We also introduce the simplified version of the model.
In section III we present a number of semi-exact results from which the
phase diagram for this case can be obtained. In section IV we 
give the results of an extensive
Monte Carlo study of this phase diagram in two dimensions. Finally,
in section V, we present our conclusions. 

\section{II. An interacting two-tolerant trail model of RNA}
RNA is a heteropolymer whose primary structure consists of a sequence of 
ribonucleotide bases. Of these there are four types, which are denoted by
the symbols C, G, U and A. The particular sequence of these bases
determines the three dimensional structure through base-base interactions,
of which pairing and stacking are the most important.

\begin{figure}
\includegraphics{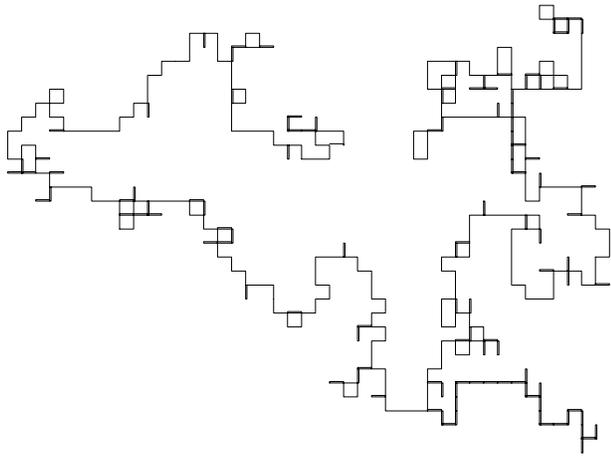}
\caption{A two-tolerant trail of $500$ steps. Doubly visited edges
are represented by double lines. \label{fig.1}}
\end{figure}

We make a lattice model for RNA starting from a two-tolerant trail \cite{MEEA}.
This is a lattice random walk that can visit each edge of the
lattice at most twice. The walk has $L$ steps, each of which corresponds
to a monomer of RNA. Doubly visited edges correspond to bonded base
pairs. The restriction to allow only doubly visited edges thus takes
into account the saturation of the base-base interactions. In Fig. \ref{fig.1}
we show a typical two-tolerant trail of 500 steps. When step $i$ and $j$
are on the same edge, we say that they are {\it bonded}. The two-tolerant
trail consists of sets of connected bonded steps (corresponding to helices
in real RNA) alternating with sets of singly visited edges (corresponding
to loops and bulges) \cite{molbiol}. With each $L$-step two-tolerant trail ${\cal {T}}$ we associate
a set $S_{\cal {T}}$ whose elements are the bonded steps of the trail:
$S_{\cal{T}} = \{(i,j):\ i,j\ \mbox{are bonded}\}$. 
In most of the existing theoretical work on RNA the following restriction
is put on the bonded steps: when both $(i,j)$ and $(i',j')$ are bonded,
one only allows the combinations $ i < j < i' < j' $ and $i < i' < j' < j$.
The situation in which $ i < i' < j < j'$ will be referred to as a 
pseudoknot, even though in the biological literature this class of
structures is further divided into several types \cite{molbiol}. In our model,
pseudoknots are treated on the same footing as the other types
of bonds.

Next, we associate
to each doubly visited edge a pairing energy $\varepsilon_{i,j}$ which
depends on the nature of the bases $i$ and $j$ present on the edge.
Below we will propose a simple form for $\varepsilon_{i,j}$ inspired
by the work of \cite{BH1} and investigate the phase diagram of the resulting
model. 

Before doing that we comment here on how other physically relevant
interactions can be introduced in a natural way in our model. 
The most important of these is the stacking energy. This
is an interaction between neigbouring bonded base pairs and
and it can be modelled by associating an energy $\sigma_{i,i+1,j-1,j}$
with each element of $S_{{\cal T}}$. The value of this stacking energy
depends on the four nucleotides forming the stack. When the
pair $(i+1,j-1)$ is not bonded we put $\sigma_{i,i+1,j-1,j}=0$.

Because of the semiflexible nature of nucleid acids,
it is also necessary to add a bending rigidity 
for which we assume the form $\kappa (\vec{n}_i - \vec{n}_{i+1})^2$
with $\kappa \geq 0$. Here $\vec{n}_i$ is the unit vector in the
direction of the $i$-th step. The total energy associated
with the two-tolerant trail ${\cal T}$ thus becomes
\begin{eqnarray}
E_{{\cal T}} =  \kappa \sum_{i=1}^{L-1} (\vec{n}_i - \vec{n}_{i+1})^2
+ \sum_{(i,j) \in S_{\cal T}} \left( \varepsilon_{i,j} +
\sigma_{i,i+1,j-1,j} \right)
\label{for.1}
\end{eqnarray}

The thermodynamic properties of the RNA-model can then be
determined from the partition sum
\begin{eqnarray}
Z_L = \sum_{{\cal T}} \exp{\left(-\beta E_{{\cal T}}\right)}
\label{for.2}
\end{eqnarray}
where the sum is over all $L$-step two-tolerant trails and $\beta$
is the inverse temperature $\beta=1/k_B T$. Moreover, if ${\cal A}$
is a particular set of bonded monomer pairs we can determine
the probability $p_{{\cal A},L}$ that this set occurs by calculating
the ratio
\begin{eqnarray}
p_{{\cal A},L} = \frac{Z_{{\cal A},L}}{Z_L}
\label{for.3}
\end{eqnarray}
where $Z_{{\cal A},L}$ is given by an expression similar to (\ref{for.2})
with the sum restricted to trails ${\cal T}$ for which ${\cal A} \subset
S_{{\cal T}}$. In this way, it is possible to calculate, for example,
the probability that a pseudoknot appears as a function of temperature.

Within this model we can calculate the thermodynamic properties of an
RNA molecule with a given sequence of bases by taking from the
literature the estimated values of the relevant pairing and stacking
energies. Clearly, such a calculation can only be performed numerically.
Here, we follow another route with the main purpose of gaining
a more detailed insight into a simplified version of our model.
Such an approach can give useful information against which the results
of a full numerical calculation can be understood. In the simplified
model we neglect the stacking energies and the bending rigidity.
Moreover, we modify the nucleotide dependence of the pairing
energies. A common approach in the physics literature is to take
it as random \cite{Pagnani,BH2,BH3,Krzakala}. In this way, the
study of RNA can be linked to that of random systems such
as spin glasses. However, clear differences between random and
real RNA have been pointed out \cite{Higgs2}. The latter has an
evolutionary evolved, correlated sequence of bases, which is
such that most often the ground state secondary structure is
less degenerated than that of random RNA. An expression for
the pairing energy which has this feature and which lends
itself to detailed analysis was introduced in \cite{BH1}.
Following that work, we take
\begin{eqnarray}
\varepsilon_{i,j} = \varepsilon_0 + \tilde{\varepsilon}\delta_{i+j,L+1}
\label{for.4}
\end{eqnarray}
(where $\delta$ is the Kronecker-delta and from now on we take
$L$ even). In the physical region of interest $\varepsilon_0<0$
and $\tilde{\varepsilon}<0$. The
second term in (\ref{for.4}) favors the formation of just one
hairpin. This structure thus corresponds to the native
state of our model. With all these simplifications, (\ref{for.1})
becomes
\begin{eqnarray}
E_{{\cal T}} = \varepsilon_0 I + \tilde{\varepsilon} N
\label{for.5}
\end{eqnarray}
where $I$ is the total number of bonded base pairs and $N$ the
total number of {\it native} interactions (i.e. those
pairs $i,j$ for which $\delta_{i+j,L+1}=1$). Finally,
we introduce $q=\exp\left(-\beta \varepsilon_0\right)$ and
$\tilde{q}=\exp\left(-\beta \tilde{\varepsilon}\right)$, so
that the partition sum (\ref{for.2}) becomes
\begin{eqnarray}
Z_L (q,\tilde{q}) = \sum_{{\cal T}} q^I \tilde{q}^N
\label{for.6}
\end{eqnarray}
In the rest of this paper we study the phase diagram of the
model defined by (\ref{for.6}).

\section{III. The phase diagram}
The two-tolerant trail was originally introduced as a simple
model for the coil-globule transition of homopolymers \cite{MD}.
The authors of that work studied the two-tolerants trails
with attractive self-interactions, which corresponds
to our model for $\tilde{q}=1$. A closer investigation \cite{Enzo}
revealed however that in the low temperature phase,
the universal properties of the trail are not those
of a collapsed globular polymer, but coincide with
those of branched polymers (BP). In the high temperature
phase it was found \cite{Enzo} that the universality class of two-tolerant
trails is that of the self-avoiding walk (SAW) \cite{Carloboek}. 
These results were based on exact enumerations in two dimensions and on a
study of the model on fractal lattices.
More recently,
we investigated the non-interacting two-tolerant trail (i.e. $q=\tilde{q}=1$)
with a Monte Carlo method that will be discussed in the
next section. We also found clear evidence that, at least in two
dimensions, the non-interacting trail has the critical
properties of the SAW. Let us therefore assume for
the moment that along the line $\tilde{q}=1$, our model
has a phase transition at some $q_c(1) > 1$ between a
SAW-phase and a branched polymer phase. A more
quantitative investigation of this transition will
be presented in the next section. We now show
how this assumption, together with a number
of other, semi-exact results, leads to a qualitative
determination of the form of the phase diagram.

For this purpose, we introduce the free energy $f(q,\tilde{q})$
which is defined as
\begin{eqnarray}
f(q,\tilde{q})= \lim_{L \to \infty} \frac{1}{L} \ln Z_L(q,\tilde{q})
\label{for.7}
\end{eqnarray}
Besides this free energy it is convenient to introduce the
connective constant $\mu_2(q,\tilde{q})=\exp\left(f(q,\tilde{q})\right)$.
The existence of the limit in (\ref{for.7}) can only be proven rigorously for
$q\leq 1,\ \tilde{q}\leq 1$ \cite{DocPet}. The proof is 
based on concatenation arguments and is a straightforward
extension of that for the SAW \cite{Madras}.
On the basis of an exact enumeration we recently determined
the estimate \cite{DocPet}
\begin{eqnarray}
\mu_2(1,1) = 3.486 \pm .003
\label{for.8}
\end{eqnarray}
on the square lattice.
Moreover, it can also be proved that in the non-interacting
case the connective constant for two-tolerant rings (i.e. two-tolerant
trails whose last step ends at the starting point) equals that
of all two-tolerant trails. 
 Again, this result can be shown
by extension of the proof of a similar result for self-avoiding
walks \cite{Madras}. For more details, we refer to \cite{DocPet}.
In the following, we will assume that (\ref{for.7}) exists.

Firstly, we investigate the behaviour of the free energy at fixed $q$ and
as a function of $\tilde{q}$. Consider therefore the limit $\tilde{q}=0$ where
the only contribution to the free energy $f(q,0)$ comes from trails
without native contacts. A special subset of these are the trails
in which the first $L/2$ steps and the last $L/2$ steps are in different halfspaces.
It is known that for quite general type of walks, the free
energy of walks limited to a halfspace is the same as that of
unrestricted walks (i.e. $f(q,1)$ in our case) up to surface corrections
\cite{polsur}. These however vanish in the
limit $L \to \infty$. Thus, the free energy of the trails without native contacts
is bounded from below by that of trails living in a half-space, which equals
that of trails without spatial restrictions. Therefore, $f(q,1) \leq f(q,0)
,\ \forall q$. However, since the free energy is a non-decreasing function
of $\tilde{q}$ we must conclude
\begin{eqnarray}
f(q,\tilde{q}) = f(q,1) =\ln \mu_2(q,1) \ \ \ \ \ \ 0\leq \tilde{q} \leq 1
\label{for.9}
\end{eqnarray}

Secondly, it is possible to obtain a lower bound to the partition sum (\ref{for.6}) by considering only the contribution
from two-tolerant trails with a maximum number of native contacts, i.e.
those for which $N=L/2$. Each of these is a walk whose first $L/2$
steps is a one-tolerant trail (i.e. a walk that can visit each edge
of the lattice at most once), and which then retraces these same
steps in reverse order during its last $L/2$ steps. For this type
of walk $I=L/2$ and therefore
\begin{eqnarray}
f(q,\tilde{q}) \geq \frac{1}{2} \left[\ln \mu_1 + \ln q + \ln \tilde{q}\right]
\label{for.10}
\end{eqnarray}
Here $\mu_1$ is the connective constant for non-interacting one-tolerant
trails. For example, on the square lattice one has $\mu_1=2.72058 \pm .00020$
\cite{LimMeir}. 

Together (\ref{for.9}) and (\ref{for.10}) imply the existence
of a phase transition at some $\tilde{q}_c(q)$. Moreover, one trivially
arrives at the bounds
\begin{eqnarray}
0 \leq \ln \tilde{q}_c (q) \leq 2 \ln \mu_2(q,1) - \ln q - \ln \mu_1
\label{for.11}
\end{eqnarray}
It also follows that for each $q$ fixed, the free energy equals the value given by
(\ref{for.9}) as long as $\tilde{q} \leq \tilde{q}_c(q)$.

\begin{figure}
\includegraphics{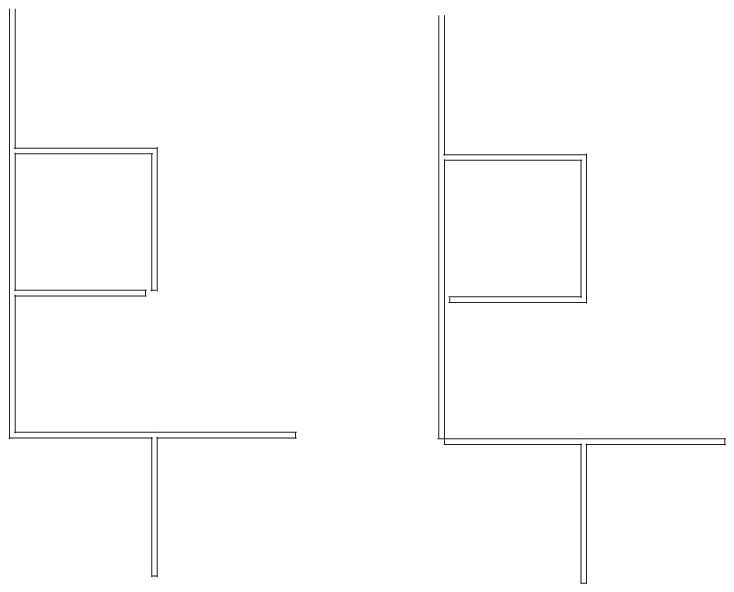}
\vspace{1cm}
\caption{Two two-tolerant trails with $I=L/2$ that can be mapped
onto the same bond lattice animal. \label{fig.2}}
\end{figure}

Further information on the upper bound in (\ref{for.11}) can be obtained
from the following reasoning. For $\tilde{q}=1$ and for $q \to \infty$, the partition sum is dominated by trails in which each
edge is visited twice. 
Each such trail of $L$ steps has the appearance of a weakly embedded bond
lattice animal \cite{Carloboek} whose $L/2$ bonds correspond with the
doubly visited edges as we show for an example in Fig. \ref{fig.2}. 
But in the same figure we show that the mapping is not one-to-one.
If we suppose that the number of trails that are mapped onto the
same lattice animal is not extensive in $L$ we can conclude that
for $q \gg 1$
\begin{eqnarray}
\ln \mu_2(q,1) \approx \frac{1}{2} \left[ \ln \mu_{BP} + \ln q \right]
\label{for.12}
\end{eqnarray}
Here, $\mu_{BP}$ is the connective constant for weakly embedded bond
lattice animals. Its value on the square lattice is estimated
as $\mu_{BP}=5.21\pm .006$ \cite{BP}. Hence, we can make the
upper bound (\ref{for.11}) more precise for $q \to \infty$. We obtain
\begin{eqnarray}
\ln \tilde{q}_c (q) \leq \ln \mu_{BP} - \ln \mu_1 \ \ \ \ q \to \infty
\label{for.13}
\end{eqnarray}

What is the physical nature of the phase transition whose existence we have just proven?
To answer this question, consider the density of native contacts $n(q,\tilde{q})$.
\begin{eqnarray}
n(q,\tilde{q}) = 2 \lim_{L \to \infty} \frac{\langle N\rangle}{L}= 2
\lim_{L \to \infty} \frac{\sum_{{\cal T}} N q^I \tilde{q}^N}{LZ(q,\tilde{q})}
\label{for.14}
\end{eqnarray}
The factor $2$ ensures that $n(q,\tilde{q})=1$ when the two-tolerant trail
consists of one single hairpin.
In terms of the free energy this density equals
\begin{eqnarray}
n(q,\tilde{q}) = 2 \frac{\partial f(q,\tilde{q})}{\partial \tilde{q}}
\label{for.15}
\end{eqnarray}
Thus, from our earlier results on the free energy we find that $n(q,\tilde{q})=0$
for $q < \tilde{q}_c(q)$ while this density becomes strictly
positive above $\tilde{q}_c(q)$. We therefore interpret
the phase transition at $\tilde{q}_c(q)$ as a denaturation transition, i.e.
a transition into (or out of) the
native state. Also notice that this transition exist for every value
of $q$.

As discussed in the beginning of this section, numerical evidence
shows that there is also a transition between a self-avoiding walk
regime and a branched polymer (BP) one along the line $\tilde{q}=1$, at
some critical value $q_c(1)$. Clearly, since the free energy
does not depend on $\tilde{q}$ for $\tilde{q}<\tilde{q}_c(q)$, the SAW-BP transition
has to be also present for some range of $\tilde{q}$-values, and moreover neither
the location of the transition point nor its critical properties
can depend on $\tilde{q}$. 

\begin{figure}
\includegraphics{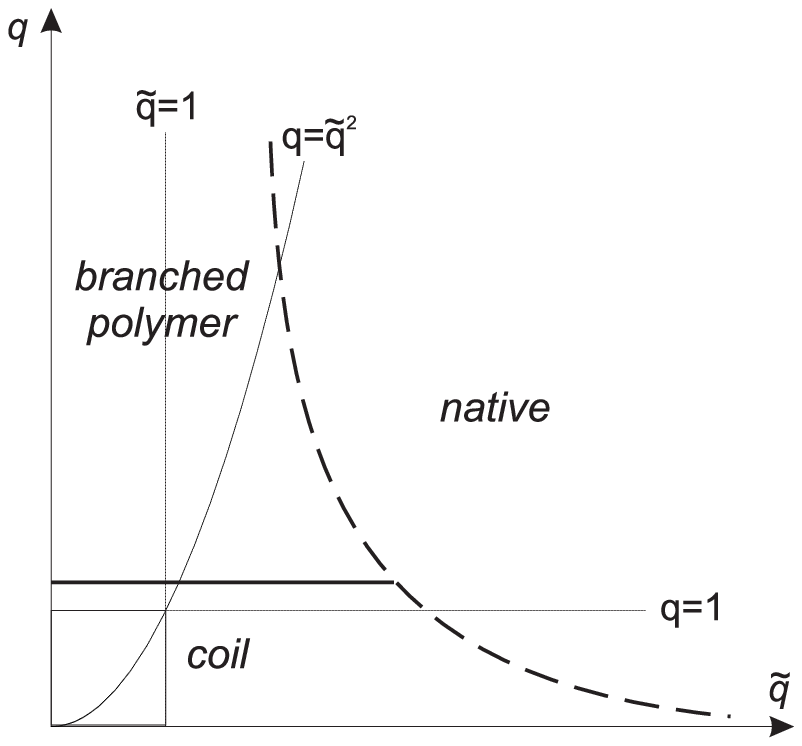}
\caption{Schematic phase diagram of our model. The thick full line is a line of second
order transitions, whereas along the broken line the transition is
first order. The thin full lines indicate the three paths along which
we investigated the model numerically. \label{fig.3}}
\end{figure}

We therefore arrive at the phase diagram shown in Fig. \ref{fig.3}.
There are three phases. At low $q$ (or $|\varepsilon_0|$) and $\tilde{q}$
(or $|\tilde{\varepsilon}|$), RNA is denaturated
and behaves as a coil in the universality class of the self-avoiding
walk. For $q>q_c(1)$, and for $\tilde{q}$ sufficiently small, there is
a collapse into a branched polymer or 'molten' phase. Finally for
$\tilde{q}$ sufficiently large, we are in the native, hairpin phase. The arguments we have given are quite
general and we therefore believe that this phase diagram is
correct, independently of dimension. 
While the denaturation transition is always present, it is possible
that the SAW-BP transition disappears above
some critical dimension. 
As an example of this we mention that in the model of reference
\cite{BH1} where pseudoknots and self-avoidance are neglected,
and which thus can be seen as a mean-field model, no evidence for the coil phase is found,
and only the transition between the native and molten phase
is present. It is interesting to remark that in that
work it was also found that in the molten phase, RNA has the
properties of (mean-field) branched polymers, as is the case here.

\section{IV. Numerical results}
In order to get a more quantitative insight into the phase diagram,
we have investigated our model in the $(q,\tilde{q})$-plane
with extensive Monte Carlo simulations. 
Further numerical insight was obtained from an exact enumeration of our model for 
$L \leq 18$. In these numerical approaches, we worked
on a square lattice.

For the simulations, we used the
pivot algoritm \cite{pivot}, a well-known technique that
generates a Markov chain in the set of all allowed walks of
a certain type. The pivot algorithm was originally introduced for
self-avoiding walks. In \cite{PCu} we investigate the extension
of this method to non-interacting two-tolerant trails,
discussing such aspects as ergodicity of the algorithm,
acceptance ratio, autocorrelation times, and so on.

Here we have to take into account interactions and do this by adding a 
standard Metropolis step to the algorithm. Moreover, we implement the 
program within a Multiple Markov Chain approach (MMC) \cite{MMC}.
In this method several Markov chains at different temperatures are run in
parallel and at regular intervals an attempt is made to
switch two trails between Markov chains at different temperature. 
Such an attempt is accepted with a probability that is a trivial
extension of that of the Metropolis algorithm.
The MMC approach has the advantage that it allows better sampling
at lower temperatures, where a standard algorithm may easily
get stuck near a metastable state. 
Besides the pivot moves, we also found it useful to add some local moves
to enhance
the performance of the algorithm.
Typically, at each temperature we performed $10^9$ Monte Carlo
steps resulting in $(.5 \sim 1.) \times 10^6$ independent configurations over
which averages are calculated.

With this approach, we investigated our model on the square
lattice along three lines. 
For these, we chose $q=1,
\ \tilde{q}=1$ and $q=\tilde{q}^2$. We now discuss 
the results.\\

\subsection{The line $q=1$.}
\begin{figure}
\includegraphics{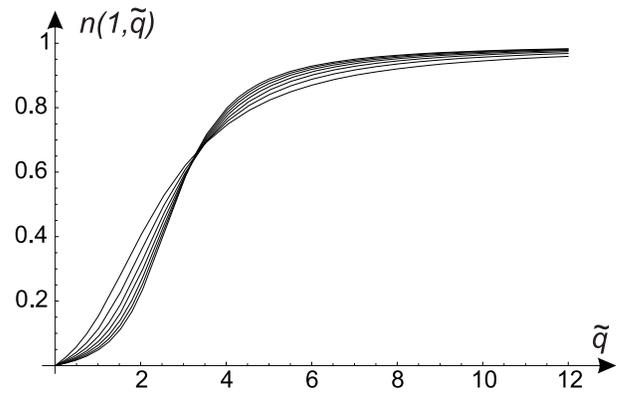}
\caption{The number of native contacts as a function of $\tilde{q}$
along the line $q=1$ for different $L$-values,
as obtained from exact enumerations. \label{fig.4}}
\end{figure}

\begin{figure}
\includegraphics{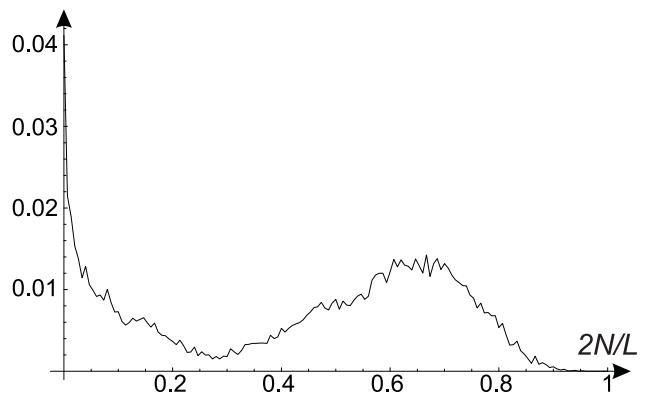}
\caption{Histogram for the density of native contacts at
the denaturation transition for $L=300$ ($q=1$). \label{fig.5}}
\end{figure}

On the basis of the phase diagram obtained in the previous section,
we expect that along this line there is only the denaturation
transition. Evidence for this transition can most easily be found
by looking at the density of native contacts. In Fig. \ref{fig.4}
we show our results for $n(1,\tilde{q})$ for two-tolerant
trails of different $L$ as obtained from the exact enumeration
data. We clearly recognise 
the behaviour predicted in section III, dressed with finite size roundings.
From the intersections of the curves for
different $L$-values a first estimate for the location
of the critical point can be made. 
Moreover, since the transition into the native state shares some
properties with the adsorption transition of a
polymer onto a surface, we expect that right at the critical
point the density of contacts scales as $n \sim L^{\varphi_n -1}$,
where $\varphi_n$ is a crossover exponent. From the exact enumeration
data shown in Fig. \ref{fig.4}, we estimate
$\varphi_n \approx .88$. However, a study of the same quantity with the Monte Carlo approach shows that this estimate
is still strongly affected by finite size effects. For example,
the value for $\varphi_n$ tends to increase to a value close
to $1$.
This suggests that the actual value
of $\varphi_n$ is one, which would be the case for a first order
transition. To verify this idea we made histograms for our
data for $n$ at different $\tilde{q}$-values. These are
indeed consistent with a first order transition.
 As an example,
we show in Fig. \ref{fig.5} such an histogram at the transition
point. There is clear evidence for two peaks, one near $n=0$,
the other around $n \approx .65$. In fact, it turns out that
from these histograms, a rather sharp estimate for $\tilde{q}_c$
can be obtained, with the result $\tilde{q}_c=4.20 \pm .02 $.

The fact that the denaturation
transition is first order can also be understood with
hindsight from a comparison with known results from DNA-models. 
Indeed, in recent years there
has been quite some interest in understanding the nature
of the denaturation transition for that biopolymer. Almost all 
existing evidence now shows that this transition is
first order, both in $d=2$ and $d=3$ \cite{Grassberger,Mukamel,
Attilio}.
When within our model, we divide the two-tolerant trail into
two halves, we can see them as the two strands of a DNA-molecule
whose starting point is halfway on the two-tolerant trail.
The native interactions, which are the only ones appearing
along the line $q=1$, can thus be interpreted as interactions
between homologous bases on the two strands of the DNA. In this
way, our model with $q=1$ becomes in a sense the dual of a recently studied
lattice model of DNA \cite{Grassberger, Attilio}, and we can therefore
expect both models to have
a similar critical behaviour.\\

\subsection{The line $\tilde{q}=1$.}
The phase transition between the coil and branched polymer regime
along the line $q=1$ is more difficult to analyse. There is
no obvious order parameter characterising this transition,
since the average number of bonded base pairs, $\langle I \rangle$,
is extensive on both sides of the transition.

We therefore investigated two other quantities. Firstly, we looked
at the specific heat, which for $\tilde{q}=1$ equals
\begin{eqnarray}
C_L (q) = \frac{1}{L} \left[\frac{\sum_{{\cal T}} I^2 q^I }
{\sum_{{\cal T}} q^I } - \left(\frac{\sum_{{\cal T}} I q^I }
{\sum_{{\cal T}} q^I }\right)^2\right]
\label{for.16}
\end{eqnarray}
While both the exact enumerations and the Monte Carlo simulations
show that the specific heat has a peak that slowly grows with
$L$, it is difficult to obtain reliable estimates for the
location of the critical point and for the crossover exponent
from these data.

\begin{figure}
\includegraphics{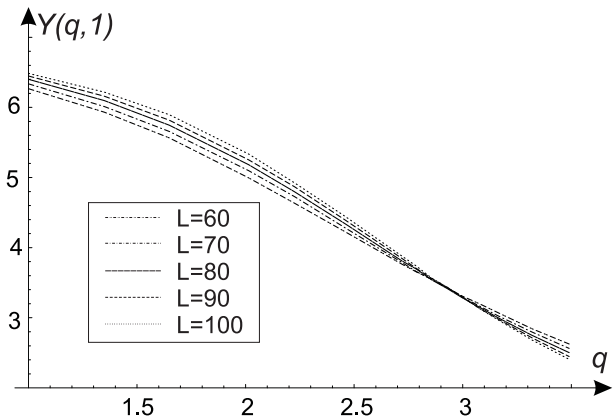}
\caption{Plot of $Y(q,1)$ (see text) for different $L$-values. \label{fig.7}}
\end{figure}

Secondly, it is possible to get information on the SAW-BP
transition from the ratio $Y(q,1)$ of the average squared end-to-end distance
over the average squared radius of gyration. It is well known that
this is a universal quantity so we expect its behaviour to be
stepwise as a function of $q$, at least for $L \to \infty$.
In fact, we expect that for large $q$ where the two-tolerant
trail visits each edge twice (see Fig. \ref{fig.2}), the end-to-end distance approaches zero and
hence $Y(q,1)$ goes to zero for large $L$. On the other hand, we verified recently that for non-interacting
two-tolerant trails $Y(1,1)=7.1235\pm .001$ \cite{PCu}, fully consistent with the
value for the SAW. Hence along the line $\tilde{q}=1$, $Y(q,1)$ should
assume this value below $q_c(1)$, and then drop to zero.
In Fig. \ref{fig.7} we present our data for $Y(q,1)$, which
have the expected behaviour, dressed with finite size rounding.
From these results we are able to obtain the most accurate
estimate for $q_c(1)$ which equals $2.91 \pm .08$.

\begin{figure}
\includegraphics{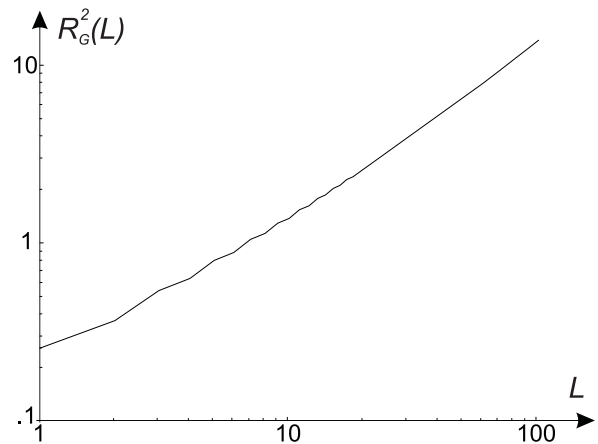}
\caption{Plot of $R_G^2(L)$ versus $L$ in the BP-phase at $q=3.49,\ \tilde{q}=1$. \label{fig.10}}
\end{figure}

In Fig. \ref{fig.10} we show data for the squared radius of gyration $R_G^2(L)$ as a
function of $L$ in the BP-phase.
From this we can obtain the geometric exponent $\nu$ since
\begin{eqnarray}
R_G^2(L) \sim L^{2\nu}
\label{for.16.1}
\end{eqnarray}
From a fit of the exact enumeration and Monte Carlo data in the BP-phase
we find $\nu \approx .55$. This is still far from the best known value
for two-dimensional branched polymers which is $\nu=.64075\pm.00015$ \cite{Stauffer}.
This difference is probably due to strong corrections to scaling. Indeed, 
this also happens 
for the non-interacting situation where very long trails can be
simulated, up to $L=7500$. From them we obtain $\nu=.749\pm.001$, as should
be expected for a walk in the SAW universality class.
However, this exponent is only recovered for $L > 200$. We expect
that the $\nu$-exponent for branched polymers will show up if one
studies longer trails at temperatures sufficiently below the transition.
But that regime is difficult to probe with our Monte Carlo technique.

\subsection{The line $q=\tilde{q}^2$.}
From the phase diagram shown in Fig. \ref{fig.3},  from the available
estimates for $q_c(1)$, and from (\ref{for.13})  we expect that along this line
two phase transitions will be encountered. The first one is the SAW-BP
transition, which can be analysed most suitably from an investigation of
the ratio $Y(q,\tilde{q})$. The BP-native transition on the other hand, should show itself
through a study of the density of native contacts.

\begin{figure}
\includegraphics{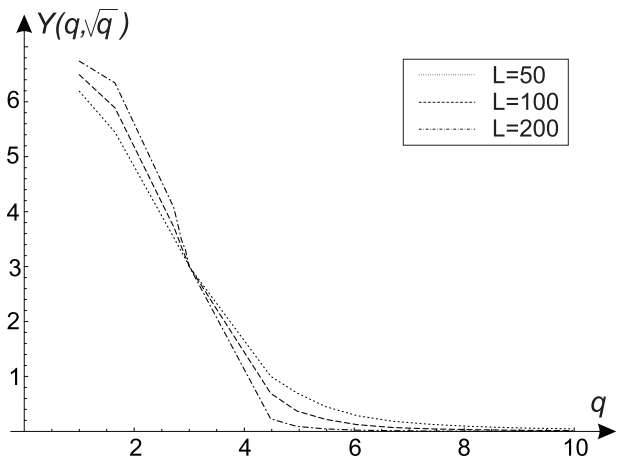}
\caption{Plot of $Y(q,\sqrt{q})$ for different $L$-values. \label{fig.8}}
\end{figure}

In Fig. \ref{fig.8} we show our results for $Y(q,\sqrt{q})$. As was the case
before, the data for different $L$-values intersect, yielding the estimate
$q_c=3.00\pm .06$. This result is within the numerical accuracy the same as that
for $\tilde{q}=1$, as was predicted in section III.

\begin{figure}
\includegraphics{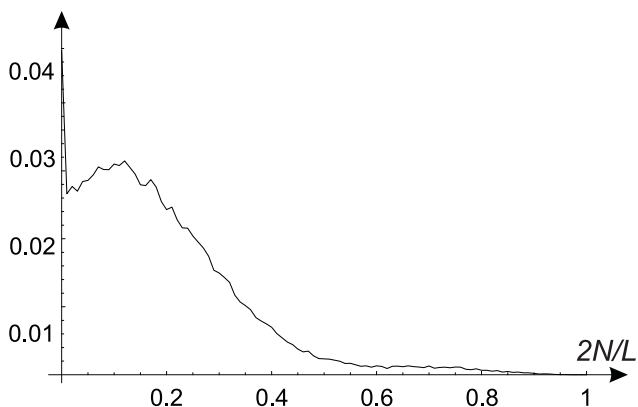}
\caption{Histogram for the number of native contacts at the BP-native
transition along the line $q=\tilde{q}^2$. 
The data are for $L=200$ and $\tilde{q}=2.17$. \label{fig.9}}
\end{figure}

From a study of the density of native contacts, we conclude that the  
transition between the
branched polymer and the native phase is also
first order. Since in this case it is a transition between two
rather dense phases, it is more difficult to obtain a reliable estimate
for $\tilde{q}_c$. We find $\tilde{q}_c=2.17\pm .15$. Fig. \ref{fig.9}
shows a histogram for the number of native contacts at this point.
 
\section{V. Conclusions}
In this paper, we have introduced a lattice model, based on a two-tolerant trail
that seems well suited to investigate thermodynamic properties of RNA. We have found
that for a simple version of the interaction energies, there are three phases
and we have investigated in detail the transitions between these phases
with a Monte Carlo method.

We believe that the structure of the phase diagram that we found here is
not particular for the choice of the interaction energy (\ref{for.4})
but would be similar also for other choices of $\varepsilon_{i,j}$ 
that break the homogeneity
of the interaction energies. Indeed, it is known that, at least in mean-field
theory \cite{BH3},  the inclusion
of a random part in the interaction energy gives rise to the appearance at
low temperature
of a  spin-glass like phase. This phase then plays the role of the native one.
It therefore seems quite possible that taking into account self-avoidance and for rather general
choices of non-homogeneous interaction energies, one recovers
the three phases found here. 

Upon lowering the temperature at fixed values of the interaction energies $\varepsilon_{i,j}$
there are thus two possible scenario's. Either one goes directly into
the native regime, or one goes through an intermediate molten phase.
We believe that the first order transition line approaches the line
$\tilde{q}=1$ when $q \to \infty$, although we could not prove this
and simulations in this regime are difficult with the pivot algorithm. If this belief is
correct it seems that the molten phase, where a description in terms
of homogeneous interactions is correct, can never be the stable one at
very low temperatures. 
 
Our results can be used as a starting point for studies which are of
more interest from the point of view of molecular biophysics. 
Firstly, still within the context of the simplified model we are investigating
the probability of occurence of pseudoknots. It is intuitively obvious that in
the hairpin-phase the occurence of pseudoknots will be severely surpressed.
But one would be interested to obtain a more quantitative insight into this
issue.

Another application for which we think that our model can be useful is
an investigation of the elastic properties of RNA. These have been measured
recently using micromanipulation techniques \cite{pulRNA}. The theoretical study of
elastic properties of biomolecules is usually performed within simple
continuum models such as the worm like chain (WLC)\cite{Marko}. This is certainly a
very good approach when effects of self-avoidance can be neglected.
For non-interacting models of polymers, it is known \cite{deGennes} that as soon as
an infinitesimal force is applied, the polymer becomes stretched
and in such a regime it can be expected that it can be described
by the WLC or in terms of directed polymers.
However, for homopolymers below the theta-point it has been established
that they undergo a transition to a stretched phase only for
forces greater then a critical force $F_c > 0$
\cite{GrassbergerHsu,Flavio}. Then, in the whole
region where the forces are below this threshold, effects of self-avoidance
are of importance. We expect that a similar scenario might hold within
the low-temperature phases (BP and native)
of RNA.  

Finally, the scenario which we have found here for the denaturation
transition could be quite general and also be of relevance for
proteins. Also in that case it is possible that depending on the
ratios of relevant interactions, the denaturation transition takes
place immediately, or through an intermediate molten phase.

{\bf Acknowledgement} We would like to thank R. Kawai for the use
of his computer cluster.

\end{document}